\documentclass[a4paper,11pt]{article}
\pdfoutput=1 

\usepackage{jheppub} 

\usepackage[T1]{fontenc} 

\usepackage{amsmath, amssymb, amsthm, graphicx, epsfig, fancyhdr,epsfig, slashed}

\usepackage{tikzsymbols}
\usepackage{natbib}
\usepackage{float}
\usepackage{tikz,xcolor,hyperref}
\usepackage{bm}
\usepackage{url}
\usepackage{braket}
\usepackage{mathtools}
\usepackage{placeins}
\usepackage{booktabs}
\usepackage{graphicx}
\usepackage{bbold}
\usepackage[normalem]{ulem}

\makeatletter \gdef\@fpheader{}\makeatother

\def\a{\alpha}

\def\G{\Gamma}

\def\D{\Delta}
\def\ve{\varepsilon}

\def\r{\rho}

\newcommand{\eq}[1]{Eq.~(\ref{#1})}
\newcommand{\fig}[1]{Fig.~\ref{#1}}

\newcommand{\beq}{\begin{equation}}
\newcommand{\eeq}{\end{equation}}

\def\SONp{\text{SO}(N+1)}
\def\SON{\text{SO}(N)}
\def\SONm{\text{SO}(N-1)}
\newcommand{\GG}{\Pi}
\def\eq#1{eq.~(\ref{#1})}
\def\sec#1{sect.~\ref{#1}}
\def\app#1{app.~(\ref{#1})}

\newcommand{\be}{\begin{equation}}
\newcommand{\ee}{\end{equation}}
\newcommand{\bea}{\begin{eqnarray}}
\newcommand{\eea}{\end{eqnarray}}
\newcommand{\bal}{\begin{aligned}}
\newcommand{\eal}{\end{aligned}}
\newcommand{\blp}{\Bigl(}
\newcommand{\brp}{\Bigr)}
\newcommand{\blb}{\Bigl[}
\newcommand{\brb}{\Bigr]}
\newcommand{\Blp}{\Biggl(}
\newcommand{\Brp}{\Biggr)}
\newcommand{\Blb}{\Biggl[}
\newcommand{\Brb}{\Biggr]}
\newcommand{\nn}{\nonumber}

\title{\boldmath Phases of Pseudo-Nambu-Goldstone Bosons}

\author[a]{Fotis Koutroulis,}
\author[b]{Matthew McCullough,}
\author[a,c,d]{Marco Merchand,}
\author[a]{Stefan Pokorski}
\author[a]{and Kazuki Sakurai }

\affiliation[a]{Institute of Theoretical Physics, Faculty of Physics, University of Warsaw, ul. Pasteura 5, 02-093 Warsaw, Poland}
\affiliation[b]{CERN, Theoretical Physics Department, Geneva 23 CH-1211, Switzerland}
\affiliation[c]{KTH Royal Institute of Technology, Department of Physics, SE-10691 Stockholm, Sweden}
\affiliation[d]{The Oskar Klein Centre for Cosmoparticle Physics, AlbaNova University Centre, SE-10691 Stockholm, Sweden.}

\emailAdd{fotis.koutroulis@fuw.edu.pl}
\emailAdd{matthew.mccullough@cern.ch}
\emailAdd{marcomm@kth.se}
\emailAdd{stefan.pokorski@fuw.edu.pl}
\emailAdd{kazuki.sakurai@fuw.edu.pl}

\preprint{CERN-TH-2023-172}

\abstract{ We study the vacuum dynamics of pseudo-Nambu-Goldstone bosons (pNGBs) for $\SONp \rightarrow \SON$ spontaneous and explicit symmetry breaking.  We determine the magnitude of explicit symmetry breaking consistent with an EFT description of the effective potential at zero and finite temperatures.
We expose and clarify novel additional vacuum transitions that can arise for generic pNGBs below the initial scale of $\SONp \rightarrow \SON$ spontaneous symmetry breaking, which may have phenomenological relevance.  In this respect, two phenomenological scenarios are analyzed: thermal and supercooled dark sector pNGBs.  In the thermal scenario the vacuum transition is first-order but very weak.  For a supercooled dark sector we find that, depending on the sign of the explicit symmetry breaking, one can have a symmetry-restoring vacuum transition $\SONm \rightarrow \SON$ which can be strongly first-order, with a detectable stochastic gravitational wave background signal. 
}

\begin{document}
\maketitle
\flushbottom


\section{Introduction}
PNGBs \cite{Nambu:1960tm,Goldstone:1961eq} arise in nature, as phonons, magnons, pions and in a broad range of theoretical scenarios.  It is no surprise that they are abundant.  It is a theorem that whenever a continuous global symmetry is spontaneously broken that NGBs will arise \cite{Goldstone:1961eq}.  Furthermore, it is widely believed that there can be no exact continuous global symmetries in nature (more precisely, in gravitational theories \cite{Hawking:1975vcx,Zeldovich:1976vq,Zeldovich:1977be,Banks:1988yz,Banks:2010zn}), in which case any NGB will, in reality, be a pNGB.  Thus, while the effective field theory (EFT) description of the low-energy behaviour of exact NGBs is an interesting object for theoretical study, it is likely that in nature the physics below the scale of spontaneous symmetry breaking is dominated by the scalar potential generated for pNGBs, since it contains the most relevant operators.

Since the structure of the pNGB potential determines the vacuum dynamics it is well-motivated to map the connections between explicit symmetry breaking sources in a UV theory and the vacuum structure and dynamics in the IR, since this aspect is physically relevant for pNGBs that are realised in nature.  Once this map is firmly established one can then determine and/or classify the plausible phases of pNGB vacua and their dynamics.

Ref.~\cite{Durieux:2021riy} established the first part of this programme for an $\SONp\to\SON$ spontaneous and explicit symmetry breaking pattern.  The fundamental building blocks of explicit symmetry breaking were found to be the irrep spurions of $\SONp$ which preserve an $\SON$ subgroup.  Each such spurion gives rise, in the IR, to a unique Gegenbauer scalar potential which is an eigenfunction of the Laplacian on the $N$-sphere.  Any general pNGB potential for $\SONp\to\SON$ can thus be decomposed as a sum of Gegenbauer polynomials.  Note that this is strongly analogous to the solution of the Hydrogen wavefunction in quantum mechanics.  The angular momentum $|j,0\rangle$ eigenstates correspond to a non-zero expectation value for the spin-$j$ irrep of $\text{SO}(3)$ which gives rise to the $j^{\text{th}}$ Legendre polynomial, which is simply an $\text{SO}(3)\to\text{SO}(2)$ Gegenbauer polynomial.  Any wavefunction which is a superposition of angular momentum eigenstates may be written as a sum of Legendre polynomials. Thus what we are familiar with for angular momentum in Hydrogen maps to the pNGBs of $\SONp\to\SON$ breaking, where the spatial rotation global symmetry becomes an internal global symmetry.

With this organisation of pNGB potentials complete the next logical step, which is to understand the vacuum dynamics, is the focus of this work.  Throughout we are concerned with the same $\SONp\to\SON$ spontaneous and explicit symmetry breaking pattern.  We focus for the most part, as a benchmark, on a single Gegenbauer pNGB potential, in the understanding that the lessons learned will map, in a straightforward way, into a sum of Gegenbauer potentials for any form of pNGB potential.

We begin by ascertaining the conditions under which the EFT description of the potential is valid, both at zero and finite temperature (specifically in the region of an interesting vacuum transition).  This effectively places a quantitative constraint on the magnitude of the explicit symmetry breaking tolerable.  Violation of this constraint implies a potential for which one does not have a controlled series expansion in the explicit symmetry breaking, whether at tree-level or at higher loop orders.

Subject to this constraint we then explore the vacuum dynamics for pNGBs, which we find to be rich and varied.  It should be noted that throughout there is explicit $\SONp\to\SON$ breaking thus, in terms of exact global symmetries, there is no formal phase transition, since only $\SON$ is an exact symmetry of the Lagrangian.  
However, since this explicit symmetry breaking is small, one does have a sense in which the fields, which play the role of order parameters, undergo vacuum transitions.

In this work we find that below the scale of spontaneous $\SONp\to\SON$ breaking, which is driven by the development of a non-zero value for the $\SONp$ radial mode, there are generically additional pNGB vacuum transitions.  There is an additional critical temperature at which the pNGBs themselves develop a vacuum expectation value, triggering a further stage of spontaneous $\SON\to\text{SO}(N-1)$ breaking. This breaking is due to the explicit symmetry breaking, but the change in order parameter is independent of the magnitude of the explicit symmetry breaking.   The reverse can also occur, with a pattern of $\SONp\to\text{SO}(N-1)$ breaking followed by a further stage of $\text{SO}(N-1)\to\SON$ symmetry restoration at lower temperatures.

It follows to determine the nature of these pNGB vacuum transitions.  There are two classes to consider, namely thermal and supercooled.  
In the thermal case we find that the transition is generically weakly first-order. On the other hand, when the pNGB sector is supercooled we find that the vacuum transition, leading to symmetry restoration, can be strong enough to generate detectable GW signatures. We finish with conclusions and future speculations.


\section{pNGB Potential Regime of Validity}
We consider an EFT containing the pNGBs $\psi$ arising from the spontaneous breaking of an approximate global symmetry at the scale $f$.  We define the action at zero temperature as
\beq\label{LEFT}
\mathcal{L} = \tfrac{1}{2}g_{ij}(\psi) \partial_\mu \psi^i \partial^\mu \psi^j + \mathcal{O}(\partial^4) - \ve V_\ve(\psi)- \ve^2 V_{\ve^2}(\psi)- \mathcal{O}(\ve^3) +... + \mathcal{L}^{\text{CT}}~~,
\eeq
where we have Taylor expanded in derivatives and in $\ve$, which is, by assumption for pNGBs, a small parameter associated with a source of explicit symmetry breaking.
$\mathcal{L}^{\text{CT}}$ represents the counterterms required for renormalisation.

Before commencing with any concrete calculations some considerations are in order concerning the validity of this EFT.  To be effective, it must be valid for some range of energies and field scales.  For the former, scattering amplitudes involving derivatives will scale as $(p^2/M^2)^j$, where $j$ is some integer and $M$ is the cutoff energy of the EFT, often associated with the mass of the radial mode of spontaneous symmetry breaking or some other UV scale such as the mass scale of intermediate vector resonances.  In any case, the EFT description breaks down, by assumption, whenever $|p^2| \sim M^2$.

Equally important is the parameter $\ve$.  In order to be considered pNGBs there must be some range of field values over which there is some sensible notion of perturbative calculability within the EFT and of a scale separation with the UV.  For pNGBs the field range is periodic in the spontaneous symmetry-breaking scale $\sim 2 \pi f$.  Due to this periodicity we will require that the EFT description is valid and affords a degree of perturbative calculability over all pNGB field values.

To determine the potential limits on the magnitude of $\ve$ it is helpful to consider the case of pions.  Were the quark masses to be comparable to the QCD scale, or the QED gauge coupling to be $e\sim 4 \pi$ in the vicinity of the QCD scale, there would be no sense in which one would have had light pions at all, as they would naturally have mass at the QCD scale.   Following this, it is tempting to diagnose EFT validity using the pNGB masses.  However, mass-scale separation alone seems insufficient.  For instance, in a scenario with two large sources of explicit symmetry breaking $\ve_1,\ve_2 \sim 1$ one could in principle fine-tune their independent contributions to a pNGB potential to give a small mass-squared in the global vacuum, generating a scale separation $m^2_\psi \ll M^2$.  However, one would have no control over perturbative corrections to the form of the pNGB potential, either at tree-level at the matching scale or in the IR at higher loops, due to the underlying magnitude of explicit symmetry breaking.  We must therefore be more pragmatic in determining the requirement on $\ve$ for the EFT description to be valid.  The condition cannot simply be that $m^2_\psi \ll M^2$, which is seemingly necessary but not sufficient.  Therefore we opt for the imprecise, but practical, condition that the pNGB potential at $\mathcal{O}(\ve)$ must be a good approximation to the full potential with all quantum corrections included.  In other words, while $\mathcal{O}(\ve^2)$ and higher terms will exist, they must not qualitatively alter the form of the pNGB potential.

The one-loop Coleman-Weinberg potential provides a useful diagnostic in this respect.  For pNGBs this is given by \cite{Binetruy:1984yx,Alonso:2015fsp,Durieux:2021riy}
\begin{equation}
V^{\rm CW} = \frac{1}{2} \mathrm{Tr} \int \frac{d^4 p}{ (2\pi)^4} \log \left[p^2 + \ve g^{-1} \left( \frac{\delta^2 V_\ve}{\delta \psi^2} - \frac{\delta V_\ve}{\delta \psi}\,\Gamma \right) \right]~~,
\end{equation}
where $\Gamma$ are the Christoffel symbols.  The field-dependent curvature (or mass-squared) entering this expression is
\begin{equation}
\mathcal{M}_\ve^2(\psi) = \ve g^{-1} \left( \frac{\delta^2 V_\ve}{\delta \psi^2} - \frac{\delta V_\ve}{\delta \psi}\,\Gamma \right) ~~,
\end{equation}
whose trace is simply the Laplace-Beltrami operator acting on the space spanned by the pNGBs.  Notably, this depends on the geometry of the manifold on which the pNGBs live.  In all of our applications we will be interested in the scenarios in which the spontaneous symmetry breaking pattern is
\begin{equation}
\frac{\SONp}{\SON} \cong \mathcal{S}^N ~~,
\end{equation}
which we recall consists of the set of points a fixed distance from the origin in $\mathcal{R}^{N+1}$.  For the sake of illustration, we focus on scenarios in which the explicit symmetry breaking follows the same pattern, preserving the $\SON$ subgroup.  As a result, we may parameterise the $N$ Goldstone bosons on this manifold through the unit vector living in $\mathcal{R}^{N+1}$ as
\beq \label{eq:nonlinear_field}
\boldsymbol{\phi} = f \sin \frac{\GG}{f} \begin{pmatrix}
           \text{n}_{1} \\
           \text{n}_{2} \\
           \vdots \\
           \text{n}_{N} \\
         \cot \frac{\GG}{f}
         \end{pmatrix}\;,
\eeq
where $\mathbf{n} \cdot \mathbf{n} = 1$.  
Thus, in this picture, $\GG/f$ essentially corresponds to the angle between the Goldstone boson direction and a given arbitrarily chosen axes in $\mathcal{R}^{N+1}$.

In these coordinates we have that the relevant mass-squared matrix is
\beq\label{M2GG}
\mathcal{M}_\ve^2(\GG) = \ve \begin{pmatrix}
           \frac{\cot(\frac{\GG}{f})}{f} V'_{\ve} \mathbb{1}_{N-1} & \mathbb{0} \\
     \mathbb{0}  &   V_{\ve}''
         \end{pmatrix}\;.
\eeq
where
\be
V'_{\ve} \equiv
\frac{ \partial V_{\ve}}{\partial \Pi}  \,\,\,~~~~ {\rm and } \,\,\,~~~~ 
 V_{\ve}''
 \equiv
\frac{\partial^2 V_{\ve}}{\partial \Pi^2}  ~~.
\ee
Thus, considering the traces of products of this matrix which will arise in perturbative calculations, it suffices to consider the Laplace-Beltrami operator
\bea
\Delta_{\mathcal{S}^N} V_\ve = V_{\ve}'' + (N-1) \cot \frac{\GG}{f} \frac{V'_{\ve}}{f}  ~~.
\label{eq:Laplacian}
\eea
As a result, truncating the momentum integral at the UV-cutoff, the zero-temperature effective potential at one-loop is
\begin{eqnarray}
V & = & 
V^{(0)}+V^{\text{CW}}+V^{\text{CT}}  \\
&  = & \ve \left[V_\ve + \frac{M^2}{32 \pi^2} \Delta_{\mathcal{S}^N} V_\ve + V^{\text{CT}}_\ve \right] + \nonumber \\
& & \ve^2 \bigg[ V_{\ve^2}  + \frac{1}{64 \pi^2}\bigg\{ \left(V_{\ve}'' \right)^2 \bigg( \log \left( \frac{\ve}{M^2}V_{\ve}''  \right) - \frac{1}{2}\bigg) \nonumber \\
&& + (N-1) \left(\frac{\cot \frac{\GG}{f}}{f} V'_{\ve} \right)^2 \bigg( \log  \left( \frac{\ve}{M^2} \frac{\cot \frac{\GG}{f}}{ f} V'_{\ve} \right) -\frac{1}{2}\bigg)  \bigg\} \nonumber \\
& & + \frac{M^2}{32 \pi^2} \Delta_{\mathcal{S}^N} V_{\ve^2}+ V^{\text{CT}}_{\ve^2} \bigg] +\mathcal{O}(\ve^3) +... \nonumber
\label{eq:effpotT0}
\end{eqnarray}
Here the terms denoted $V^{\text{CT}}$ represent the counterterms required to renormalise the pNGB potential and $V^{(0)}$ is the tree-level scalar potential.  Thus we see that if $\Delta_{\mathcal{S}^N} V_\ve$ has a very different functional form to $V_\ve$, the counterterm potential cannot be similar in form to $V_\ve$, implying some level of fine-tuning between UV/threshold corrections, which must exist, and the bare potential in order to realise the form of $V_\ve$.  If, however, they are of a similar functional form then the $\mathcal{O}(\ve)$ corrections will not destabilise the pNGB potential at that order. We will return to this possibility in due course.

More immediately relevant is that the $\mathcal{O}(\ve^2)$ effective potential corrections may significantly modify the qualitative nature of the potential.  This would signify the breakdown of the effective description of the pNGB potential.  Thus we will only work with EFTs for the pNGBs in which $\ve$ is sufficiently small that the physics of the zero-temperature potential is well described at leading order in $\ve$, hence
\beq\label{VGGT0}
V  \approx  \ve \left(V_\ve + \frac{M^2}{32 \pi^2} \Delta_{\mathcal{S}^N} V_\ve + V^{\text{CT}}_\ve \right) ~~,
\eeq
is a reasonable approximation to the pNGB potential at zero temperature.  This can only be diagnosed on a case-by-case basis, and so we leave further discussion of this aspect until a specific model has been chosen.

Now moving to finite temperature and following by analogy with the Coleman-Weinberg potential, under the same set of assumptions, the full finite-temperature potential at one-loop is, to a leading approximation,
\begin{equation}
V(T)  =  
V^{(0)}+V^{\text{CW}}+V^{\text{CT}} + V^{\text{T}} ~~,
\end{equation}
where \cite{Dolan:1973qd}
\begin{eqnarray}
V^{\text{T}} & = & \frac{T^4}{2 \pi^2} \mathrm{Tr}  J_B \left( \frac{\mathcal{M}^2(\GG) }{T^2} \right)  ~~,\\
& = & \frac{T^4}{2 \pi^2} \left( J_B \left( \frac{\ve V_\ve'' }{T^2} \right) + (N-1) J_B \left( \frac{\ve \cot \left( \frac{\Pi}{f} \right) V_\ve' }{f T^2} \right) \right),
\end{eqnarray}
and the function $J_B$ is
\begin{equation}
J_B (x) =  \int^\infty_0 dy y^2 \log\left( 1-\exp^{-\sqrt{y^2+x}} \right)  ~~.
\label{eq:JB}
\end{equation}
Since we now have a new energy scale in the theory, $T$, we ought to reconsider the conditions under which one has an appropriate description of the physics.  For $T\to 0$ we have that $V^{\text{T}} \to 0$, as expected, thus at very low temperatures we may simply use the zero-temperature effective potential already described.

At high temperatures we may also perform an expansion, in which case
\begin{equation}\label{VTM}
V^{\text{T}} \approx - N\frac{\pi^2}{90} T^4 +\ve \frac{T^2}{24} \Delta_{\mathcal{S}^N} V_\ve -\frac{T}{12 \pi} \left(\ve V_\ve''  \right)^{3/2} - (N-1)\frac{T}{12 \pi} \left(\ve \cot \frac{\GG}{f} \frac{V'_{\ve}}{f} \right)^{3/2} +...~~.
\end{equation}
The validity of this expansion rests on two separate aspects.  The first is that the high-temperature expansion should be convergent, hence when the system lies at high enough temperatures we require that the physics is, to a good approximation, described by the second term alone, with the third remaining a subleading correction.  The second aspect concerns the non-analyticity of the $J_B$ function, and hence of the third term of \eq{VTM}.  This non-analyticity generates imaginary terms in the effective potential in regions where $ \partial^2 V^{(0)}(\GG)/ \partial \GG^2 < 0$.  Since the effective potential is, by definition, a real scalar quantity this signals a breakdown in the effective description of the physics.

Without committing to a specific model in which one can calculate the magnitude of the various terms this is as far as we may proceed, thus we now commit to a specific class of scenarios.

\section{Gegenbauer Goldstones}
Experience with many physical systems, including electrostatics and thermodynamics, suggests that when one encounters the Laplacian the natural functions to work with are the eigenfunctions, satisfying an equation of the form $ \Delta_{\mathcal{S}^N} V_\ve(\GG) \propto V_\ve(\GG)$.  This is an eigenfunction problem and the solutions which are analytic in $\GG$ are the well-known Gegenbauer polynomials \cite{Durieux:2021riy}
\beq
\Delta_{\mathcal{S}^N} G^{\frac{N-1}{2}}_n (\cos \GG/f) = - \frac{n (n + N-1)}{f^2} G^{\frac{N-1}{2}}_n (\cos \GG/f)~~,
\eeq
where the eigenvalues and eigenfunctions are characterised by the two integers, $N \ge 1$ and $n \ge 0$.
In the application to the pNGB potential,
these integers are related to the 
explicit symmetry breaking pattern $\SONp\to\SON$ realised by a symmetry-breaking spurion in the $n$-index symmetric irrep of $\SONp$ \cite{Durieux:2021riy}.

\begin{figure}[!t]
\centering
\includegraphics[scale=.8]{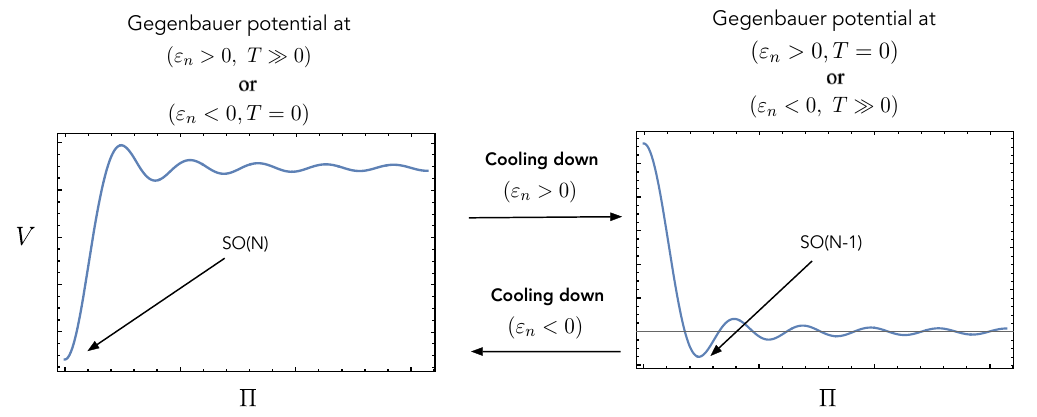}
\caption{\it A cartoon picture showing the functional form of the Gegenbauer thermal effective potential given by \eq{eq:highT}, for the temperature asymptotics $T = 0 $ and $T \gg 0$, when the symmetry-breaking parameter, $\ve_n$, is either positive or negative. The high-temperature limit terminates below the radial mode mass $M$, otherwise the original, approximate, symmetry is restored and the effective description of the model in terms of pNGBs is lost.
Cooling down will lead to $SO(N)$ symmetry restoration or breaking depending on the sign of $\ve_n$.
}
\label{Combined}
\end{figure}

Motivated by this we will thus consider a zero-temperature pNGB potential of the form
\bea
\label{VGGtree}
V(\GG,0) & \approx& \ve_n  V_{\ve_n} +\mathcal{O}(\ve^2) \nonumber \\
& \approx& \ve_n f^2 M^2 G^{\frac{N-1}{2}}_n (\cos \GG/f) + \mathcal{O}(\ve^2) +... ~~.
\eea 
where note that from now on $\ve$ would carry the subscript $n$ to distinguish the above choice from the general pNGB case of \eq{VGGT0}. No summation over the index $n$ is implied. 
The typical shape of the Gegenbauer potential at zero temperature ($T=0$) is shown in the left ($\ve_n < 0$) and right ($\ve_n > 0$) panels of \fig{Combined}.
Note that for positive $\ve_n$ the global minimum is at a scale $\langle \GG \rangle \sim 5.1 f/n$ \cite{Durieux:2021riy}, whereas for negative $\ve_n$ the global minimum is at the origin.
Importantly, this potential is radiatively stable, since at leading order in this spurion only this term can arise irrespective of the UV physics.
Since any general potential may be constructed from a linear sum of Gegenbauer polynomials the lessons learnt from studying the single polynomial case will, in generic cases, extend to more general pNGB potentials that can arise for the $\SONp\to\SON$ case.
\subsection{pNGB Potentials at Zero Temperature}\label{pNGBPZT}

With this model we may now return to our general requirement of \eq{VGGT0}.  We consider the zero-temperature potential at one-loop
\bea
V^{(1)}(\GG,0)&\approx& \ve_n  \left[ \left(1 -  \frac{ n(n + (N-1)) M^2}{32\pi^2 f^2} \right)  V_{\ve_n} + V_{\ve_n}^{\rm CT} \right] \nn\\
 &&+ \ve_n^2\left[ V_{\ve_n^2} +  V_{\ve_n^2}^{\rm CT}  - \frac{1}{128\pi^2} \left( \blp V_{\ve_n}''(\Pi) \brp^2 + (N - 1) \frac{\cot^2 \frac{\Pi}{f}}{f^2} \blp V_{\ve_n}'(\Pi) \brp^2 \right) +... \right]~~,\nn\\
\eea
where the ellipses denote the logarithmic terms.  We see that at $\mathcal{O}( \ve_n)$ the quadratic divergence may be absorbed into a counterterm of the same functional form as the initial potential, reflecting the radiative stability of this potential.  However, we also see that, regardless of the form of the potential at $\mathcal{O}( \ve_n^2)$, there are calculable terms proportional to $\ve_n^2$.  In order for the EFT to be valid it is necessary that these terms are subdominant to the leading one. 

Since it is the point at which the second derivative of the potential is maximal in magnitude, to establish the maximal permitted value of $\ve_n$ we now focus our discussion around the origin of field space. The Gegenbauer potential and its derivatives scale there as
\bea\label{GGscaling3}
V_{\ve_n}(0) &=&  f^2M^2 \frac{(n + N -2)!}{n!(N - 2)!} ~, \nn\\
V_{\ve_n}''(\Pi)\Big |_{\Pi = 0} &=& \cot \frac{\Pi}{f} \,  \frac{V_{\ve_n}'(\Pi)}{f}\Big |_{\Pi = 0} =  - M^2 (N - 1) \frac{(n + N -1)!}{(n-1)!N!} ~~.
\eea
Thus we find the condition 
\bea
\frac{\ve_n^2}{128\pi^2 } N \blp V_{\ve_n}''(0) \brp^2 &\ll& \ve_n V_{\ve_n}(0) ~, 
\eea
which, under \eq{GGscaling3}, is reduced to  
\be\label{MMc23}
|\ve_n| \ll 
128 \pi^2\, 
\frac{f^2}{M^2}  \frac{N! (n-1)!}{(n+N-1)! n (N-1) (n+N-1)} \equiv \ve^0_{n,\rm max } ~,
\ee
as a necessary condition for the EFT expansion to be valid at zero temperature, hence the upper-script $0$ in $\ve_{n, \rm max}$ refers to the zero-temperature case.
\subsection{pNGB Potentials at Finite Temperature}\label{pNGBPFT}

After renormalization, for this class of potentials the high (enough) temperature form 
is approximately 
\beq
V(\GG,T) \approx \ve_n f^2 M^2 \left(1 - \frac{n (n + N-1)}{24} \frac{T^2}{f^2} \right) G^{\frac{N-1}{2}}_n (\cos \GG/f) + \mathcal{O}(\ve^2) +... ~~.
\label{eq:highT}
\eeq
Thus, for temperatures satisfying
\beq
T^2 \gtrsim T_F^2 = \frac{24}{n (n + N-1)} f^2 ~~,
\eeq
where we refer to $T_F$ as the ``Flipping Temperature'', the overall sign of the scalar potential has changed, indicating a transition in the position of the global minimum relative to the zero-temperature potential, see \fig{Combined}. The functional form of the scalar potential remains unchanged up to the overall factor.  
We must, however, determine whether we may trust the EFT expansion at this temperature by checking the magnitude of the next term in the finite-temperature expansion.

We proceed as for the zero-temperature case, but now using the thermal potential in \eq{VTM}. The effective potential becomes 
\bea\label{fullVGGeff}
V(\GG,T) &\approx& -N\frac{\pi^2T^4}{90} + \ve_n \blb 1 - \frac{n(n + N -1)T^2 }{24 f^2} \brb V_{\ve_n}(\GG)  \nn\\
&-& \frac{T \, (\ve_n V''_{\ve_n}(\GG))^{\frac{3}{2}}}{12\pi} -  (N-1)\frac{T \, (\cot{\frac{\Pi}{f}}\, \ve_n V'_{\ve_n}(\GG))^{\frac{3}{2}}}{12\pi f^{3/2}}   + {\cal O}({\ve_n}^2) ~~.
\eea
Focusing around the origin of the field space and noting that the second derivative of the Gegenbauer polynomial is negative there,
the relevant constraint reads
\be\label{DVbV32}
\Big |\frac{T^2}{24} \ve_n \Delta_{\mathcal{S}^N} V_{\ve_n}(0) \Big | \gg \Big| N\frac{T}{12 \pi} \left(\ve_n V_{\ve_n}''(0)\right)^{3/2}\Big| ~ .
\ee
This is a necessary condition
for the validity of the EFT expansion at a given temperature. For $T \approx T_F$  we get
\beq
|\ve_n| \ll 6 \pi^2 \frac{f^2}{M^2} \frac{N! (n-1)!}{(n+N-1)! n (N-1) (n+N-1)} \equiv \ve^{T_F}_{n,\rm max } ~~.
\label{eq:EFTvalidity}
\eeq
This is a stronger bound than  at zero temperature, since
\beq
\ve^{T_F}_{n,\rm max } =\frac{3}{64} \ve^0_{n,\rm max } ~~. 
\eeq
The condition \eq{DVbV32} is necessary for validity at any temperature but not sufficient.
A stronger bound is obtained for 
$T=T_{\rm Crit}$, the `Critical Temperature', at which the vacuum transition is initiated.
In general $T_{\rm Crit} > T_F$, with the former defined as the temperature where the potential energy of the two relevant phases becomes degenerate (or the two phases have equal free energy density)
\be\label{F0FPi3}
V(0,T_{\rm Crit}) = V(\braket{\Pi},T_{\rm Crit}) ~ ,
\ee
where $\braket{\Pi}$ is the pNGB value at the degenerate vacuum.  From \fig{Combined} note that no matter which cooling-down picture we consider, the potential admits one global minimum around the field-space origin justifying our choice of $V(0,T_{\rm Crit})$ as the free energy of one of the degenerate phases. 

Using the effective potential of \eq{fullVGGeff}, assuming for now $\ve_n > 0$, the above equality gives
\be
T_{\rm Crit}^2 + [B_\ve T^2_F] \frac{T_{\rm Crit}}{f} - T_F^2 = 0 ~~,
\ee
with the solution
\be\label{TcrTFl3}
T_{\rm Crit} = \frac{1}{2} \Blb - B_\ve + \sqrt{\frac{4f^2}{T_F^2} + B^2_\ve} \Brb \frac{T^2_F}{f} ~~ .
\ee
$B_\ve$ is a dimensionless parameter defined as
\bea\label{BenN3}
 B_\ve &=&\frac{f \D V_{\ve,3/2}}{12\pi \D V_\ve} 
  \approx \frac{f}{T_F}\Bigg \{ \frac{T_F N \big(\ve_n V_{\ve_n}''(0)\big)^{\frac{3}{2}}}{12\pi\, \ve_n V_{\ve_n}(0) } \Bigg\} - \frac{f \big(\ve_n V_{\ve_n}''(\braket{\Pi})\big)^{\frac{3}{2}}}{12\pi \ve_n V_{\ve_n}(0) }
\eea
where we have defined
\be\label{DeltaV3}
\D V_{\ve,3/2} = N\big(\ve_n V_{\ve_n}''(0)\big)^{\frac{3}{2}} - \big(\ve_n V_{\ve_n}''(\braket{\Pi})\big)^{\frac{3}{2}} ~~,
\ee
and
\be
\D V_\ve = \ve_n V_{\ve_n}(0) \Bigg( 1 -  \frac{V_{\ve_n}(\braket{\Pi})}{ V_{\ve_n}(0)} \Bigg) \approx \ve_n V_{\ve_n}(0) > 0 ~~.
\ee
The notion of $T_{\rm Crit}$ and the validity of the EFT breaks down if $B_\ve$ has large  imaginary part.
Note that the term included in $\{ \cdots \}$ above, which is purely imaginary, has been used 
in \eq{DVbV32} to derive the  $\ve$ bound of \eq{eq:EFTvalidity}. However, that bound is not sufficient to make the left hand side of \eq{F0FPi3} (and as a consequence $B_\ve$)
to a good approximation real.
It is found that only for an $|\ve_n|$ which is at least $\mathcal{O} (10^{-2}) $ smaller than $\ve^{T_F}_{n,\rm max }$ the $\frac{f}{T_F} \{ \cdot \cdot \cdot \}$ term can safely be neglected from $B_\ve$ and
the latter then becomes 
\be\label{Beexp13}
 B_\ve \approx - \frac{f \big(\ve_n V_{\ve}''(\braket{\Pi})\big)^{\frac{3}{2}}}{12\pi \ve_n V_{\ve_n}(0)} < 0  ~ ~~ {\rm and }~ ~~  |B_\ve| \ll 1 ~~,
\ee
and is real so we can safely evaluate the critical temperature.
This stronger bound is used in this paper as the sufficient condition for the validity of the EFT in the whole relevant range of temperatures. Under that condition we obtain that $T_{\rm Crit} \gtrsim T_F$ within a few percent. The two temperatures are sometimes identified in our qualitative discussion but kept distinct in the numerical calculations.   

\begin{figure}
\centering
\includegraphics[scale=.45]{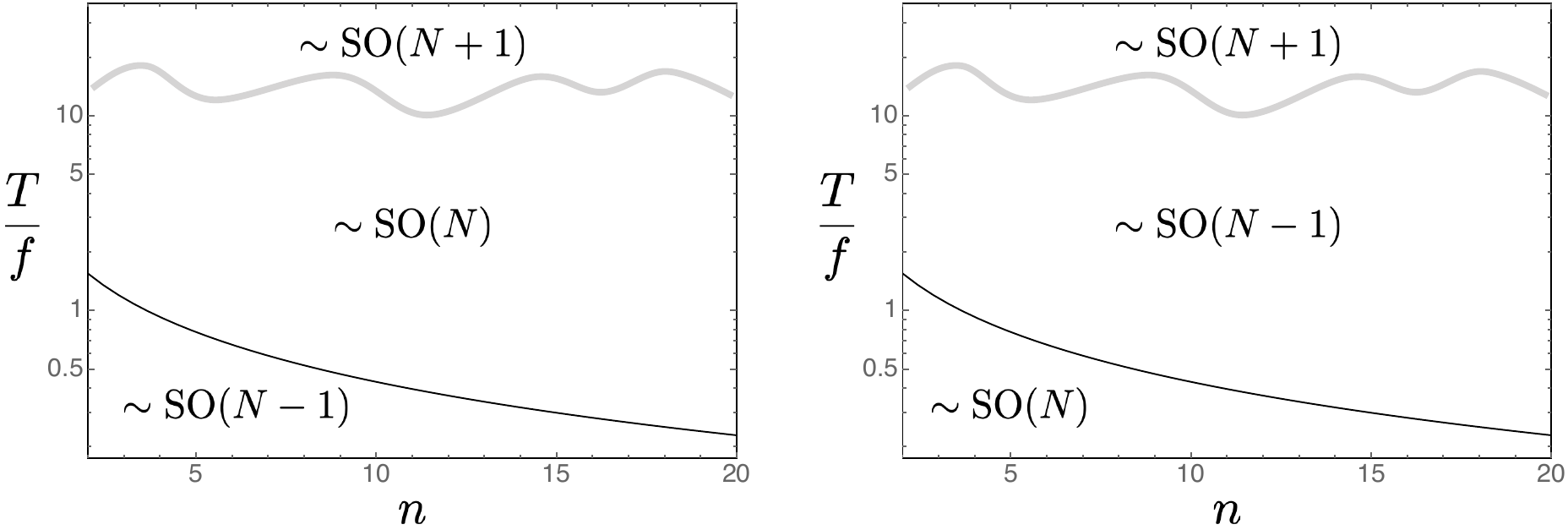}
\caption{ \it Schematic phase diagram for radiatively and thermally stable pNGB potentials, for $\ve_n >0$ (left) and $\ve_n<0$ (right).  Throughout there is explicit breaking $\SONp\to\SON$.  At high temperatures, above the mass of the radial mode, an approximate $\SONp$ is restored.  For $\ve_n >0$ at lower temperatures, $\SONp$ is spontaneously broken and at some lower temperature the exact $\SON$ is also spontaneously broken.  Whereas for $\ve_n <0$ at lower temperatures, $\SONp$ is spontaneously broken to $\SONm$ and at some lower temperature the exact $\SON$ is restored.}
\label{fig:schematic}
\end{figure}

To summarise, we see that for this class of pNGB potentials there are hierarchies of vacuum transitions.  Starting from zero temperature as the temperature is raised there will be a vacuum transition in the vicinity of the flipping temperature.  Depending on the sign of the spurion this will be from zero pNGB vev to a non-vanishing one, with $\langle \GG \rangle \propto f/n$, or vice-versa.  The nature of this transition is not yet clear from this analysis, yet its existence is clear.  Going to even higher temperatures, above the mass scale of the radial mode in the UV completion the standard symmetry-restoring transition occurs.  These scenarios are illustrated in \fig{fig:schematic}.

It is surprising and rather non-trivial that for a single spontaneous symmetry breaking scenario, with a single explicit symmetry-breaking spurion in a symmetric irrep one has a hierarchy of vacuum transitions at hierarchical scales.  It remains to determine the nature of this new vacuum transition.

\section{Cosmological Gegenbauer Phases}\label{CGP}
Having outlined the general phase structure of pNGB potentials it remains to determine any potential observable consequences of the additional pNGB vacuum transitions.  We consider a dark sector (DS) containing pNGBs with two initial conditions after the end of inflation; thermal and supercooled, however in both cases colder than the visible sector.  Given the natural origins and ubiquity of light pNGBs in quantum field theories, and given the clear evidence for the existence of dark matter, a DS scenario is well motivated and plausible.  In both cases we also investigate potential stochastic GW Background signatures arising from the vacuum transitions.

\subsection{Hot Dark Sector}
\label{sec:hotds}

We assume that the early universe dynamics is governed by the inflaton which, at the end of inflation, starts to oscillate about the minimum of its potential thus, due to its coupling to the Standard Model fields, the universe enters the reheating period.
At the same time we  consider a DS of pNGBs which is completely decoupled from (or may have an extremely small coupling to) the SM, such that it will not thermalize with the SM fields. The DS temperature, $T_h$, could be above or below the visible one, $T_v$, depending on how strongly each sector couples to the inflaton.
The ratio of temperatures after reheating,  $\xi_{\rm DS} = T_h/T_v$, is heavily constrained by Big Bang Nucleosynthesis (BBN) and Cosmic Microwave Background (CMB) measurements \cite{Planck:2018vyg, Yeh:2022heq}.

As noted, we assume $\xi_{\rm DS}<1$. This type of scenario has been investigated in  \cite{Breitbach:2018ddu, Fairbairn}. The case of $\xi_{\rm DS}>1$ is more delicate since it requires an out-of-equilibrium mechanism to inject entropy back into the SM before BBN, see e.g. \cite{Ertas:2021xeh}. For model-independent studies regarding the constraints on DS vacuum transition parameters see also \cite{PhysRevD.105.095015, Bringmann:2023opz}.\footnote{Here we will not deal with the case where $\xi_{\rm DS} = 1$, which could happen either by thermalization of the DS  with the SM thermal bath or due to specific initial conditions where the inflaton couples democratically to both sectors. We escape the former by assuming the DS has a negligible interaction or never comes into contact with SM and the latter by considering a different evolution of the two sectors during reheating.} 

A general investigation of the nature of the transition is challenging and essentially beyond the reach of standard computations.  However, subject to the requirement of small enough $\ve_n$, discussed in the previous section,
we may have some control in the vicinity of the flipping temperature. 

To proceed let us recall that the scalar potential in the DS is a Gegenbauer polynomial. The vacuum structure of such a potential is non-trivial given that different local minima coexist for a wide range of temperatures (see \fig{Combined}). Analysing its thermal history in the following, a vacuum transition is expected to occur. Particularly, for $\ve_n>0$, $\Pi$ obtains a non-zero vacuum expectation value and spontaneously breaks the SO(N) symmetry.

Before getting into a description of the phase transition details let us present an analytic estimate for the transition strength $\a$, assuming it takes place around $T\approx T_F$.  To quantify $\a$ 
we use the latent heat released normalized to the radiation energy density, which can be written as
\be\label{alphaT}
\a(T) \equiv  \frac{1}{\r_{R}}\left( \Delta V(\GG,T) - \frac{T}{4} \Delta \frac{\partial V(\GG,T)}{\partial T} \right)~, 
\ee
where the difference between the false and true vacuum is taken. The energy density is 
\bea \label{rho_total}
\r_{R} &=& \frac{\pi^2\, g_{ \Pi}^* T_h^4}{30} + \frac{\pi^2\, g_{\rm SM}^{*}(T_v) T_v^4}{30} \nn\\
 &=& \frac{\pi^2\, T_h^4}{30}\Big( N + \frac{g^*_{\rm SM}(T_v) }{\xi_{DS}^4} \Big) 
\eea
Since we consider a phase transition within the DS, the Hubble rate and the other relevant parameters are functions of $T_h$. We keep $T_v$ as a fixed initial parameter
and the number of degrees of freedom in the DS corresponds to the number of pNGBs, i.e., $g_{ \Pi}^*=N$. We evaluate the radiation degrees of freedom of the SM, $g_{ \rm SM}^*$, from tabulated data in \cite{Saikawa:2018rcs} and we keep them constant for temperatures in the vicinity of the phase transition.

Making use of the high-temperature expansion we have that the potential energy difference between false and true vacua is
\bea
\Delta V(\GG,T) &\approx& V(0,T) - V(\braket{\Pi},T)  =  \left[ 1 - \frac{T^2 }{T_F^2} \right] \D V_\ve - \frac{T\, \D V_{\ve,3/2}}{12\pi }~,
\eea
while the partial derivative with respect to temperature becomes
\bea
\frac{T}{4} \Delta \frac{\partial V(\GG,T)}{\partial T} &=& \frac{T}{4} \left[ \frac{\partial V(\GG,T)}{\partial T} \Big |_{\Pi = 0} - \frac{\partial V(\GG,T)}{\partial T} \Big |_{\Pi = \braket{\Pi}}  \right] \nn \\
 &\approx& - \frac{1}{2} \frac{T^2 }{T_F^2} \D V_\ve - \frac{1}{4}\frac{T\, \D V_{\ve,3/2}}{12\pi }~,
\eea
thus
\bea
\a(T) &\approx& \frac{\D V_\ve}{\r_{R}} \Blp \left[ 1 - \frac{1}{2}\frac{T^2 }{T_F^2} \right]  - \frac{3}{4}\frac{T\, \D V_{\ve,3/2}}{12\pi  \D V_\ve} \Brp ~.
\eea
Focusing around $T_F$, which is used as a proxy for the nucleation temperature $T_n$ since we have verified they are very close numerically, the second term in the above equation reduces to the $\{ \cdot\cdot\cdot \}$ term of \eq{BenN3} which, as follows from the discussion above \eq{Beexp13},
has to be very small for the validity of the EFT.  Thus, the transition strength becomes
\be
\a(T_F) \approx \frac{\D V_\ve}{\r_R} \left[ 1 - \frac{1}{2}\frac{T_F^2 }{T_F^2} \right] = \frac{\D V_\ve}{2\r_R} ~~.
\ee
By setting $\ve_n = 10^{-2} \ve^{T_F}_{n,\rm max }$ we obtain
\be\label{aTcriteTng}
\a(T_F) \lesssim \frac{ 0.002 }{\left( 1    + \frac{g_{\rm SM}^*(T_v)}{\xi_{\rm DS}^4 N}      \right)}~~.
\ee
The phase transition is weak because of the strong upper bound on $\ve_n$, which also controls the magnitude of the explicit breaking of the original symmetry.  This value of $\alpha$ corresponds to, at most, a very weakly first-order transition and suppressed gravitatonal wave spectrum.

Since the phase transition occurs at finite temperature under the presence of a non-negligible thermal plasma formed out of a system of pNGBs, the expanding bubble walls transmit a substantial energy density and pressure to the surrounding plasma.
Hence, the dominant source of GW production is the motion of the plasma itself, expressed in the form of sound waves. 
As described in greater detail in an \app{app:quantGW}, for the GW spectrum, under the assumption of small $\alpha$, the peak of the spectrum is
\cite{Hindmarsh:2013xza,Hindmarsh:2015qta,Hindmarsh:2017gnf,Hindmarsh:2020hop}
\be\label{Omegah2}
\Omega_{\rm sw}(\text{Peak})h^2 \approx 4 \times 10^{-7} \, \left(R_* H_*\right)^2 \left(\kappa_{\rm sw} \,\alpha \right)^\frac32 \,,
\ee
where $\kappa_{\rm sw}$ encodes kinetic energy normalized to vacuum energy. We evaluate the efficiency factor $\kappa_{\rm sw}$ using the numerical fits of \cite{Espinosa:2010hh}. $R_*$ is the average bubble size at collision.  As described in \app{app:quantGW}, we find that numerically, at the time of the transition, one has $R_* H_* \sim 10^{-6}$.  Hence we expect at most to have a spectral peak of magnitude
\be\label{Omegah3}
\Omega_{\rm sw}(\text{Peak})h^2 \lesssim 4 \times 10^{-23} \, ,
\ee
well below the expected reach of future gravitational wave detectors.  

\subsection{Supercooled Dark Sector}\label{SHS}

Let us now explore the extreme possibility that our pNGB DS is supercoooled, parameterized as $\xi_{\rm DS} \approx 0$. This may occur if, for instance, the
DS is very weakly coupled to the inflaton.
We also discuss the role of $\ve_n$'s sign. In the previous section we have assumed that $\ve_n > 0 $.
However, in principle, $\ve_n$ can be either positive or negative and, as we explain in the following, the choice of sign impacts the cosmology of the DS.

It is possible that the expansion rate of the universe is initially much faster than the bubble nucleation rate in a supercooled DS.\footnote{Since the DS is almost decoupled from the SM it will evolve independently, so we consider that the visible sector is ``frozen'' to a given temperature $T_v$.} 
As a consequence the DS can enter a period of supercooling, remaining in a local minimum until quantum tunneling towards another local or a global minimum takes place.


For $\ve_n \gtrsim 0 $ the vacuum dynamics of a supercooled DS is governed by the zero-temperature potential of \eq{VGGtree}.
\begin{figure}[!htbp]
\centering
\includegraphics[scale=.7]{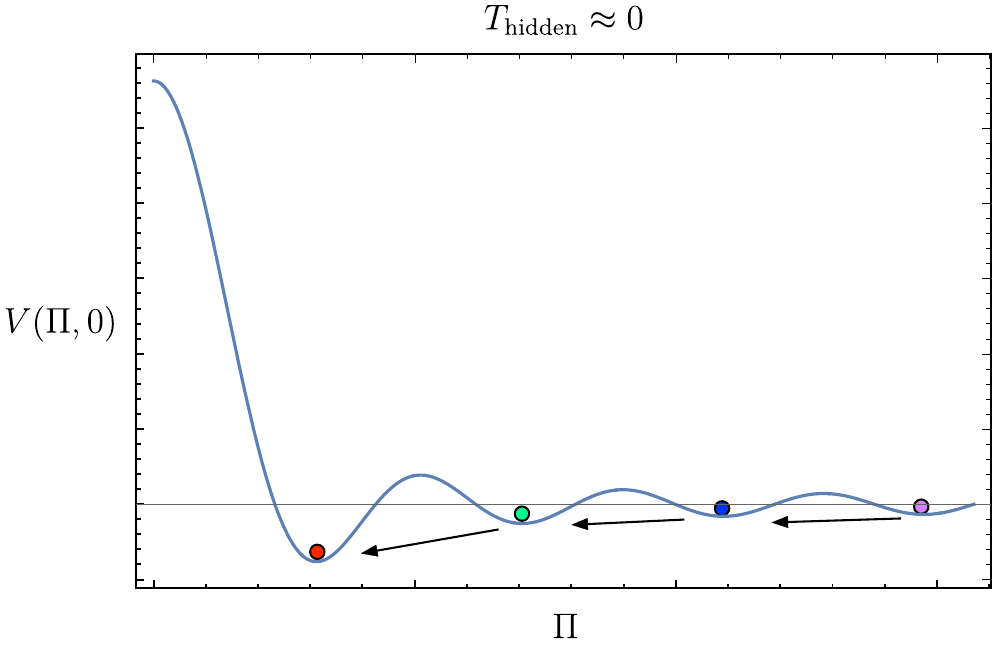}
\caption{\it Successive tunneling towards the true vacuum for the benchmark scenario $n = 15, N =4$. The colouring shows that we move from a higher $\braket{\Pi}$ ({\color{violet} purple} dot) down to smaller values until the DS reaches the deepest minimum ({\color{red} red} dot). 
}
\label{Th0PTplot}
\end{figure}
Such a scenario has interesting phenomenology as the associated potential possesses various local minima and as a consequence the supercooled DS could in principle exhibit successive vacuum transitions, depicted on \fig{Th0PTplot}, via tunneling. 
For an indicative example we consider the case when the DS is initially in the minimum depicted by the {\color{violet} purple} dot in \fig{Th0PTplot} with associated vev $\braket{\Pi_{\rm purple}}$. We calculate the probability of tunneling towards its nearest neighbor {\color{blue} blue} dot with associated vev $\braket{\Pi_{\rm blue}}$.
For this transition it is clear that 
the barrier between the vacua is large compared to the energy difference between them, therefore the thin wall approximation \cite{Coleman4} is a well motivated analytic approach.
According to this approximation and following \cite{Fairbairn}, the probability of nucleating a critical bubble via quantum tunneling is 
\be\label{Gamma4}
\G_4 = A_4 ~ e^{-S_4} \equiv  \frac{1}{R_0^{4}} \left( \frac{S_4}{2\pi} \right)^2 e^{-S_4} 
\ee
where $S_4$ is the $O(4)$-symmetric bounce solution and $R_0$ is the size of the nucleating bubble.

Moreover, following the cosine-like approximation to the Gegenbauer potential provided in Eq.\ (2.12) of \cite{Durieux:2021riy}, and employing the triangle approximation to the cosine potential, for which an anlytic expression was derived in \cite{Duncan:1992ai}, in the thin wall approximation the bounce action $S_4$ scales as
\bea\label{O4bs}
S_4 \approx  \frac{32 \pi^2}{3} \frac{(\Delta V_{\text{Max}}(\GG))^2 (\D\GG)^4}{(\Delta V(\GG))^3} ~~,
\eea
where $\D\GG $ is the leading order change in vev between vacua, $\Delta V(\GG)$ is the change in vacuum energy between the two vacua and $\Delta V_{\text{Max}}(\GG)$ is the change in vacuum energy between the vacuum and the top of the barrier between them.

The resulting expression for the bounce, in the large $n$ limit, is
\be
S_4 \sim \frac{2^{3-n-N} n^2 \pi^5 \Gamma(n+N)}{3 (N-1)^4 \Gamma\left(\frac{n+1}{2} \right)  \Gamma\left(\frac{N}{2} \right) \Gamma\left(\frac{n+N-1}{2} \right)} \times \frac{\ve^0_{n, \rm max}} {\ve_n} ~~,
\ee
which ultimately scales proportional to $n!/((n/2)!)^2$, quickly becoming very large for large $n$.  We also have that 
\be\label{R0}
R_0^4 \approx \frac{S_4}{\pi^2 \D V(\GG)} 
\ee
so substituting the above relations back to \eq{Gamma4} it becomes clear that for $\ve_{n, \rm max}^0/\ve_n$ satisfying the criteria for a controlled EFT expansion the exponential becomes extremely small.
The condition for a successful completion of the vacuum transition is
\be\label{G4H4m}
\G_4 \gtrsim H^4 ~~,
\ee
which is difficult to fulfil.  In conclusion, if the DS is for some reason localized at the {\color{violet} purple} dot then it will face an extremely slow decay rate, compared with the expansion of the universe, such that it will never completely tunnel to the {\color{blue} blue} dot in a time scale which is relevant, leading to an eternally-inflating DS. 

\begin{figure}[!t]
\centering
\includegraphics[scale=.5]{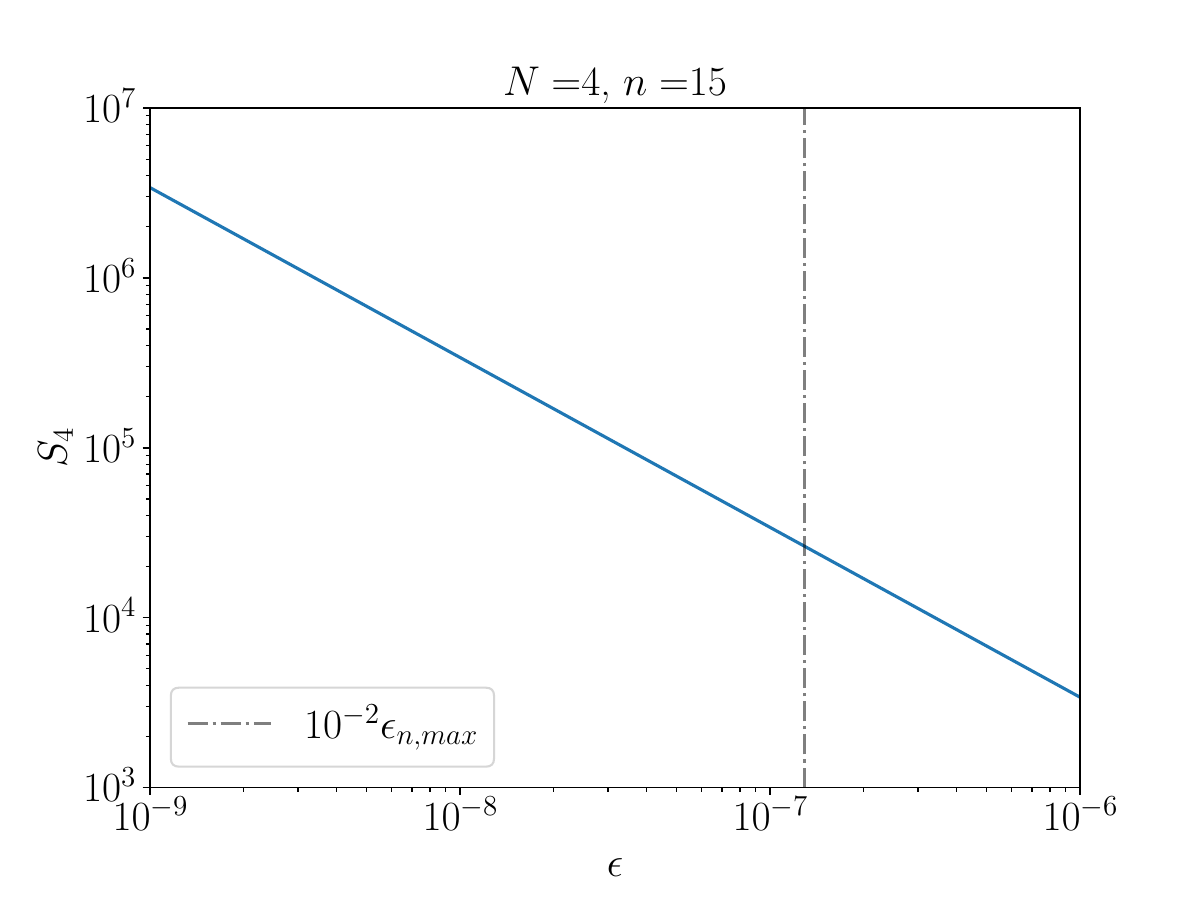}
\caption{\it The bounce solution $S_4$ evaluated numerically as a function of $\ve$ for the {\color{green} green} dot $\to$ {\color{red} red} dot transition as they are represented in \fig{Th0PTplot}. 
}
\label{S4_numerical}
\end{figure}

Naturally one is led to consider the other tunneling possibilities.  Na\"ively for transitions closer to the true global minimum one, such as the {\color{green} green} to {\color{red} red} dot vacuum transition (see \fig{Th0PTplot}), one does not expect a dramatic change since the difference in vacuum energy and the height of the barrier grow in a correlated manner, however \eq{O4bs} suggests that the change in vacuum energy may ultimately dominate such that faster tunnelling may be possible.
In such transitions the energy difference is comparable to the barrier height, hence the thin wall approximation cannot be trusted and a numerical analysis of the bounce action is required.  To this end we rely again on a modified version of CosmoTransitions \cite{Wainwright:2011kj} code.  The numerical analysis of the bounce solution as a function of $\ve_n$, for the benchmark scenario studied here, is shown in \fig{S4_numerical}, demonstrating that only a case of a large $\ve_n$, well above the upper value for an effective description of the pNGB potential, admits values of $S_4$ which could allow the vacuum transition to complete.

To conclude, we find that a supercooled vacuum transition in a DS with a single Gegenbauer potential and $\ve_n > 0 $, is highly unlikely to successfully complete unless $\ve_n$ violates the EFT bound, in which case calculability is called into question.


\subsection*{PT from a flipped potential}
\begin{figure}[!t]
\centering
\includegraphics[scale=.5]{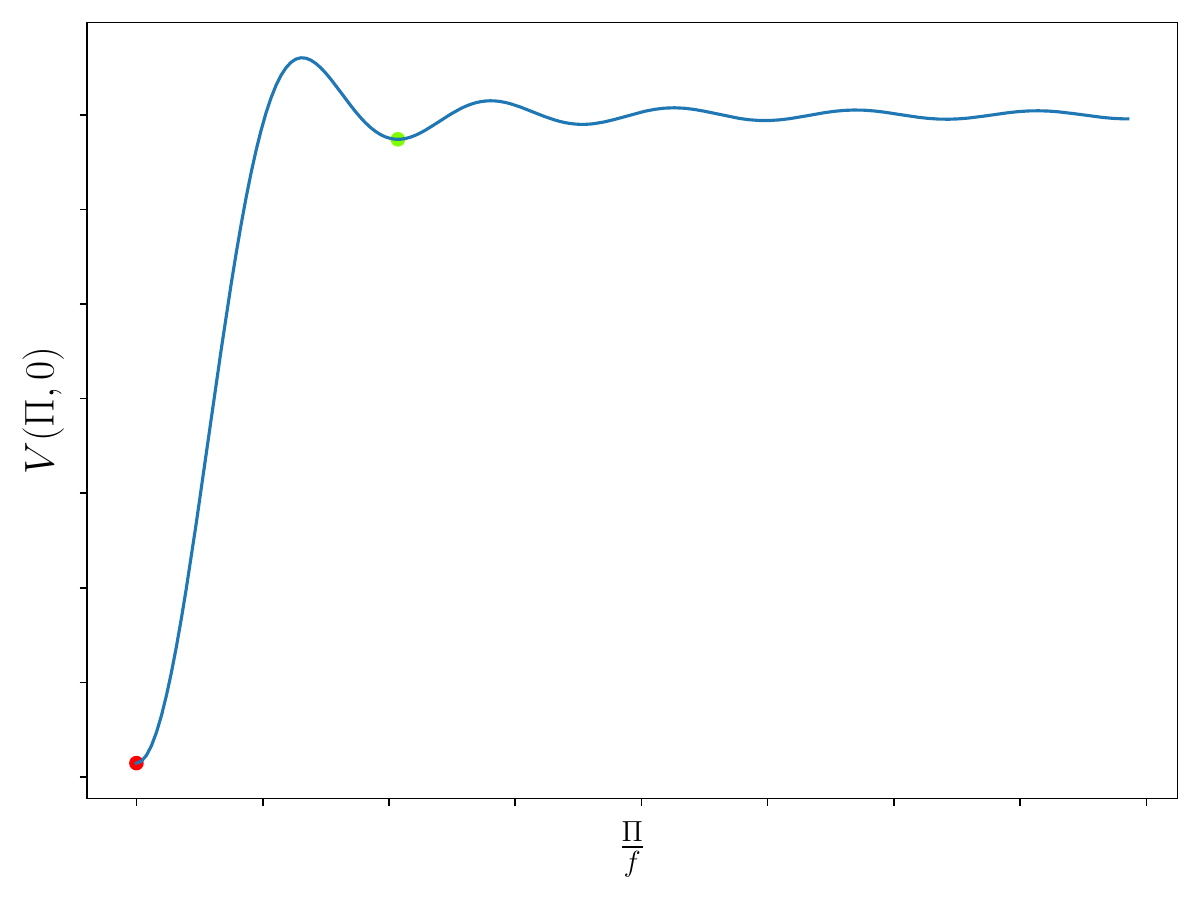}
\caption{ \it Inverted tree-level Gegenbauer potential.  With the transition from the green dot to the red dot considered. 
}
\label{flipped_pot}
\end{figure}

Now consider the case with  $\ve_n <0 $, as displayed in \fig{flipped_pot}.  We focus on the transition from the second minimum to the origin.  Notice that this process corresponds to a symmetry-restoring phase transition since the pNGB order parameter $\Pi$ has a zero vev in the true vacuum.  This transition is outside the validity of the thin-wall approximation thus we compute the constant decay rate, \eq{Gamma4}, numerically. To estimate the bubble radius at nucleation, $R_0$, we use the value at which the field profile function is halfway between the two minima.

The Hubble rate is written as  
\begin{equation}
    H^2 \equiv \frac{\pi^2 g^{*}_{SM}(T_v) T_v^4}{90 M_{\text{Pl}}^2} + \frac{\Delta V(\Pi,0)}{3 M_{\rm Pl}^2} ~,
    \label{eq:Hubble}
\end{equation}
where the first term comes from the standard radiation degrees of freedom. The second term above is the vacuum contribution and we have assumed that the DS temperature remains negligibly small. For simplicity, we fix the value of $\varepsilon_n= 10^{-2} \ve^0_{n,\rm max}$ and the resonance mass scale to $M= 4\pi f$. Thus only $N$, $n$ and the symmetry breaking scale $f$ are free parameters. 

\begin{figure}[!thbp]
\centering
\includegraphics[scale=.45]{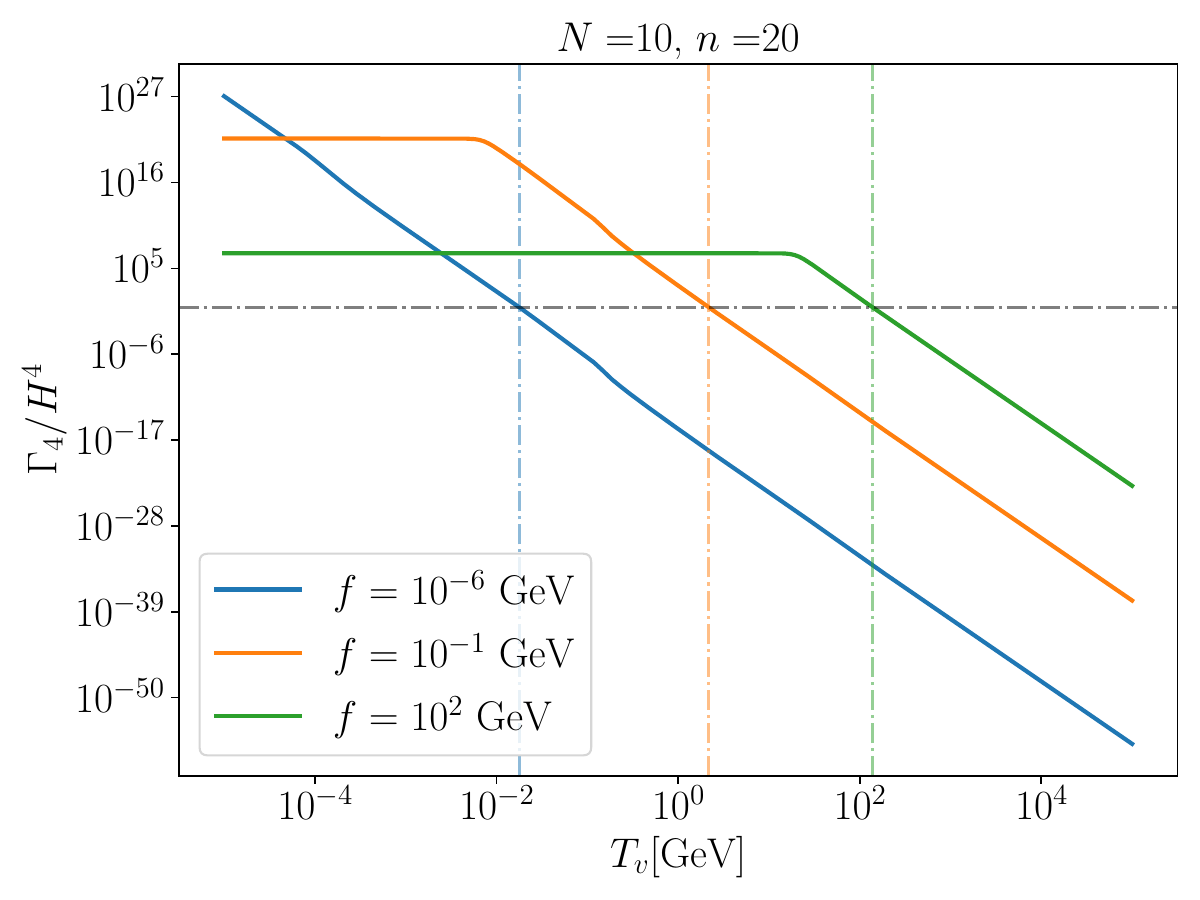}
\caption{ 
\it Ratio of nucleation rate to Hubble volume as a function of visible sector temperature for different values of the compositeness scale. The horizontal line marks the nucleation condition while the vertical lines help visualize the intersection point.   At high temperatures $\Gamma_4/H^4 \propto 1/T_v^8$ while as the temperature drops the vacuum contribution begins dominating the Hubble rate and $\Gamma_4/H^4 \approx \text{const}$.}
\label{DS_nucleation}
\end{figure}

The tunnelling rate $\Gamma_4$ is independent of the visible sector temperature and instead  all the temperature dependence is encoded in \eq{eq:Hubble}. We also find that the polynomial order $n$ has a negligible impact on the decay rate. Once one fixes $N$, $n$ and $f$, one has that $\Gamma_4/H^4 \propto 1/T_v^8$ for large $T_v$. As the temperature drops the vacuum contribution starts dominating the Hubble rate and $\Gamma_4/H^4 \approx \text{const}$. This behavior is displayed in \fig{DS_nucleation} for $N=10$ and $n = 20$  and several values of symmetry breaking scale $f$.  One can observe from this figure that the nucleation temperature is directly proportional to the compositeness scale $f$, as expected on dimensional grounds.
Notice that if a transition is too slow to occur at $T_v=0$ then it cannot start for any $T_v$.  In addition, since the potential is effectively temperature-independent, the strength parameter of the phase transition is approximately
\begin{equation}
  \alpha(T_v) \approx \frac{\Delta V(\Pi,0)}{\rho_R} ~.
\end{equation}

\begin{figure}[!tbp]
\centering
\includegraphics[scale=.45]{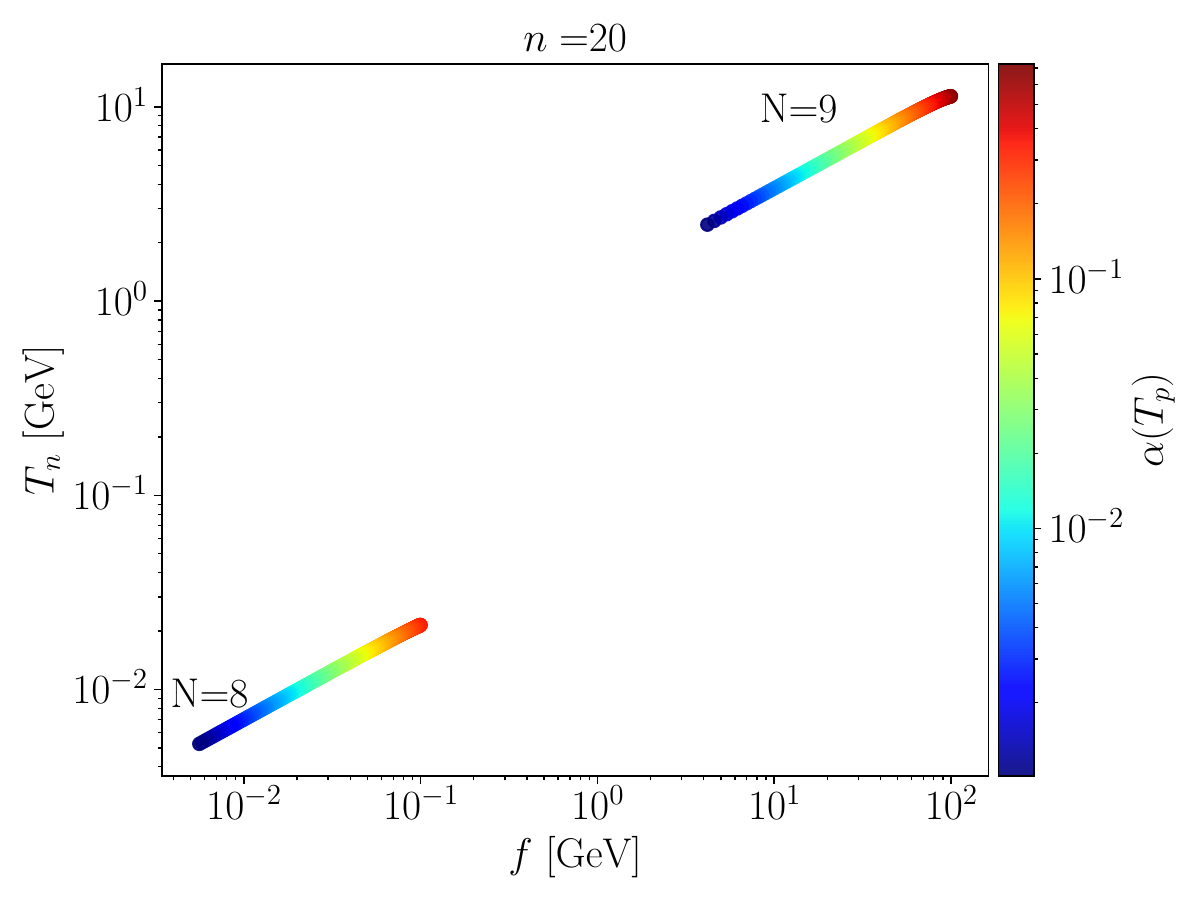}
\caption{ 
\it Nucleation temperature as a function of symmetry breaking scale $f$ with the colorbar displaying the strength parameter $\alpha$ at the percolation temperature.}
\label{alpha_plot}
\end{figure}

In \fig{alpha_plot} we show this transition strength (colorbar) alongside the behavior of the nucleation temperature as a function of symmetry breaking scale for two benchmark values of $N$. The number of pNGBs, $N$, significantly impacts the possible range of nucleation temperature due to the fact that, in our chosen parametrization, $N$ affects the barrier height and thus, through the bounce action, impacts the tunneling rate exponentially. The lines terminate at the symmetry breaking scale $f$ for which the nucleation rate matches the minimum value $\Gamma_4 \approx H^4$, as can be inferred from \fig{DS_nucleation}.  Close to this point, the nucleation condition becomes numerically ambiguous.  For smaller values of $f$ the lines are truncated at values with extremely weak vacuum transitions.  It can be observed that the strongest phase transitions are associated with the largest possible symmetry breaking scale and can attain values $\alpha \approx \mathcal{O}(1)$.

For very strong phase transitions the latent heat released accelerates the wall to relativistic velocities and the effects of the thermal plasma are suppressed. Thus the DS plasma of pNGBs exerts negligible friction on the wall and one has $v_w \approx 1$. In this case the GW signal is sourced by the collision of the walls and not by the sound waves, thus the treatment differs from \sec{sec:hotds}.   To estimate the time scale of the transition we consider the bubble number density, which for a constant decay rate reads \cite{Cutting:2018tjt}\footnote{In this expression, the gamma function $\Gamma(x)$ should not be confused with the decay rate $\Gamma_4$.}
\begin{equation}
\frac{1}{R_{*}^{3}}= \frac{1}{4}\left( \frac{\Gamma_4}{v_w}     \right)^{3/4} \Gamma\left(\frac{1}{4} \right) \left(  \frac{3}{\pi} \right)^{1/4} = \frac{1}{8 \pi} \frac{\beta^3}{v_w^3}.
\end{equation}
The GW spectrum from bubble collisions is estimated as \cite{Bringmann:2023opz}
\begin{equation}
\Omega_{\rm GW}(f)h^2 = \tilde{\Omega} \times S\left(\frac{f_g}{f_{\text{col}}}  \right),     
\end{equation}
\begin{figure}[!thbp]
\centering
\includegraphics[scale=.5]{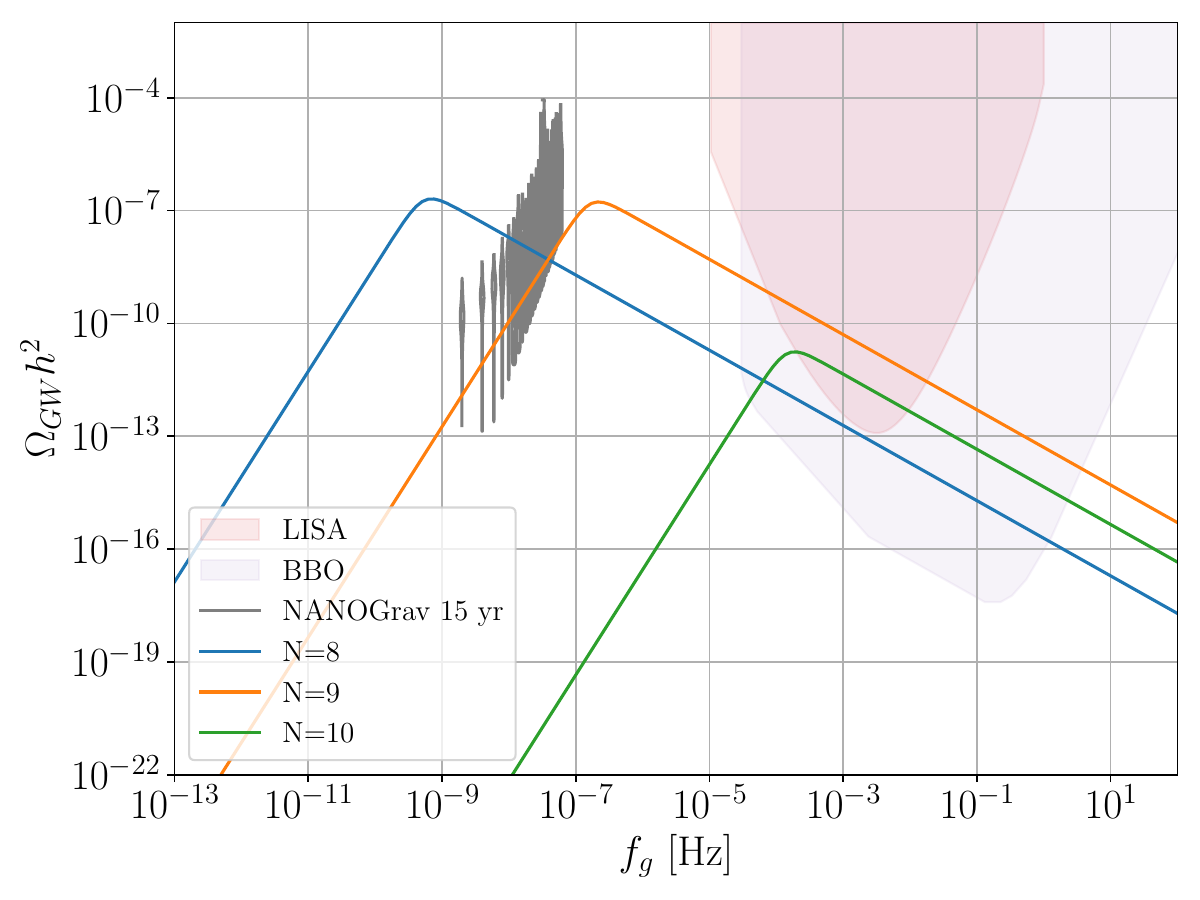}
\caption{ \it GW spectrum from bubble collisions for the strongest signals found. The red contours are the violin curves for the NANOGrav $15$ yr data obtained from \cite{NANOGrav:2023hvm} using the public tool \cite{Rohatgi2022}. The integrated sensitivity curves for LISA and BBO were obtained using \cite{Thrane:2013oya}.
}
\label{GW_results}
\end{figure}
where we write the amplitude in terms of mean bubble separation as 
\begin{equation}
\tilde{\Omega} \approx  1.7 \times 10^{-5}\ \tilde{\Omega}_{\text{bw}} (H_{\rm min}R_{*})^2 (8\pi)^{-2/3} \left( \frac{ \kappa_{\phi} \alpha(T_p)}{1+ 
 \alpha(T_p)}\right)^2 \left(  \frac{g_{*}(T_p)}{100} \right)^{-1/3},
\end{equation}
where $H_{\rm min}^2 = \Delta V/3 M_{\rm Pl}^2$, the coefficient $\kappa_{\phi}$ is obtained from the detonation approximation from \cite{Espinosa:2010hh} and the spectral function is given by
\begin{equation}
    S(x) = \frac{19 x^{14/5}}{5+14 x^{19/5}} ~.
\end{equation}
After red-shifting the peak amplitude we have that 
\begin{equation}
    f_{\text{col}}  = 1.7 \times 10^{-5} 
    (R_{*}H_{\text{min}})^{-1} (8 \pi)^{1/3} \left(  \frac{T_p}{100 \ \text{GeV}}\right) \left(  \frac{g_{*}(T_p)}{100 }\right)^{1/6}  \left(  
 \frac{f_{\text{peak}}}{\beta} \right) ~~\text{Hz},
\end{equation}
with  $ f_{\text{peak}}/\beta\approx 0.2$, $\tilde{\Omega}_{\rm bw} \approx 0.08$. In the expressions above we have used a slightly more precise percolation temperature, at which the probability to find a region of space-time still in the false vacuum has decreased to about $P(T_p )\sim e^{-1}$.

We show, in \fig{GW_results}, the predicted GW spectrum from bubble collisions for three benchmark values of $N$ where in each case we select the value of $f$ which maximises the strength of the phase transition. We display the sensitivities of the future detectors LISA  \cite{LISACosmologyWorkingGroup:2022jok,Caprini:2019pxz} and BBO \cite{Corbin:2005ny}.  As we can observe from this figure, the case $N=9$ could potentially explain the recently observed common-red spectrum from the NANOGrav $15$ yr data \cite{NANOGrav:2023hvm} which is shown as the gray curves.

\section{Summary and conclusions}
The vacuum structure and dynamics of theories possessing pNGB fields in the IR is of theoretical interest and physical importance.  Indeed, the vacuum structure of QCD itself is a rich subject rendered tractable by studying the vacuum structure of the pNGBs \cite{Witten:1980sp,Creutz:1995wf,Smilga:1998dh,March-Russell:2020lkq}.  In this work we have explored a complementary facet of pNGB vacua which arises if explicit symmetry breaking occurs due to a spurion in a non-minimal representation.  Here, again, there are metastable vacua, however they exist for different field values, as described in \cite{Durieux:2021riy}.  In this work, we have focused on the same $\SONp\to\SON$ symmetry breaking pattern and investigated the resulting vacuum dynamics, which are found to be much richer than one might na\"ively expect.

Our main result is that the `primary' phase transition associated with spontaneous $\SONp\to\SON$ breaking when the radial mode obtains a vacuum expectation value is not the end of the story.  Below this scale the pNGBs will typically undergo additional vacuum transitions unless the sources of explicit symmetry breaking take the most minimal form.

These vacuum transitions may occur in two ways.  Thermally, there is a second critical temperature scale, the `Flipping Temperature', which scales proportional to $T_F \propto f/n$ and can thus naturally be well below the spontaneous symmetry breaking scale $f$.  Crucially, at this temperature the functional form of the pNGB potential remains the same, to leading order in the spurion.  However, the overall sign flips, such that the higher temperature minimum becomes the lower temperature maximum, and vice-versa for the higher temperature maximum.  As a result, in the vicinity of the flipping temperature an additional vacuum transition occurs.  We find this is likely weakly first-order, at least for parameters consistent with a controlled EFT.

The second possibility arises non-thermally, if the pNGB sector becomes supercooled in a metastable state, which is not implausible given the existence of $\sim n$ different metastable vacua.  In this case multiple vacuum transitions can occur, with the most likely being to a nearest neighbour.  As the field approaches the global minimum the final vacuum transition can be strong enough to generate observable GWs.

The vacuum structure of our universe is of prime importance and interest in physics.  It determines the ultimate fate of the observable universe and may carry lessons about the deep UV and quantum gravity itself \cite{Obied:2018sgi}.  Spontaneous symmetry breaking is ubiquitous in nature, for which Nambu-Goldstone bosons are the physical manifestation of the vacuum structure.  Similarly, pNGBs manifest, through their vacuum structure, patterns of explicit symmetry breaking.  As a result, physically relevant lessons concerning the vacuum structure and cosmological dynamics of nature may be learned by studying pNGBs, perhaps even the case in which the Higgs boson is a pNGB; a case we leave to further study.

\section*{Acknowledgements}
The authors would like to thank Marek Lewicki and Andreas Mantziris for useful discussions.
The research of F.K., M.M., S.P. and K.S. leading to these results has received 
funding from the Norwegian Financial Mechanism for years 2014-2021, 
grant nr DEC-2019/34/H/ST2/00707. M.M. also acknowledges support from 
the Polish National Science Center grant 2018/31/D/ST2/02048.  K.S. is partially supported by the National Science Centre, Poland, under research grant 2017/26/E/ST2/00135.

\appendix

\section{Hot Sector Calculations}
\label{app:quantGW}
We now detail a numerical investigation of the phase transition for a hot DS. The theory of the vacuum decay from a local false minima to the true global minima at zero and finite temperature has been studied extensively \cite{Coleman:1977py,Callan:1977pt,Coleman:1980aw,Linde:1980tt,Linde:1981zj}. When the temperature is non-negligible the transition proceeds through thermal fluctuations by the nucleation of true vacuum bubbles within the space filled with false vacuum energy. The probability of decay per unit time and volume is
\be\label{Gamma3T}
\Gamma_3(T) =\left(  \frac{S_3(T)}{2\pi T}\right)^{3/2} T^4 e^{-S_3(T)/T}~,
\ee
where  $S_3(T)/T$ is the finite temperature Euclidean action of our pNGB model and is less than the zero-temperature one, $S_4$, around $T_F$.

The true vacuum bubble nucleates when the decay rate becomes comparable to the expansion rate of the universe. Namely, we define the bubble nucleation temperature by
\be
\Gamma_3 \approx H^4 \big |_{T \equiv T_n}~,\label{nuc_condition}
\ee
where the Hubble rate is given by
\be\label{H2full}
H^2 =  \frac{ \rho_R }{3  M_{\text{Pl}}^2}   + \frac{\Delta V(\GG,T)}{3 M_{\text{Pl}}^2} ~, 
\ee
which includes the contribution from the potential energy difference between false and true minima and $M_{\text{Pl}}=2.4\times 10^{18}$ GeV is the reduced Planck mass.  

As mentioned earlier, the hidden and visible sectors have independent temperatures and cool at different rates.  From \eq{rho_total} above we can read off the total effective number of degrees of freedom as
\begin{equation}
    g_* = \left( N + \frac{g^*_{\rm SM}(T_v) }{\xi_{DS}^4}  \right) ~.
\label{g_star_total}
\end{equation}
The time scale of the transition is given by
\be 
\frac{\beta}{H} \equiv T \frac{d}{dT} \left(  \frac{S_3(T)}{T} \right)\bigg{|}_{T \to T_n}\,.
\label{eq:betaHN}
\ee 
To compute the action we solve the equation of motion for the system, also known as the bounce solution.  This can be considerably simplified by considering the parametrization of \eq{eq:nonlinear_field} and allowing for a vev only in the $\GG$ direction such that
\be
 \Box \Pi -   \frac{\partial V(\GG,T)}{ \partial \Pi}=0 ~~.
\ee
We use a modified version of the publicly available code CosmoTransitions \cite{Wainwright:2011kj} to compute the Euclidean action.

Finally, it is necessary to have an estimate for the bubble wall velocity. This requires an out-of-equilibrium computation of the deviation from equilibrium of all the particle distribution functions. While this is still a very active area of research \cite{Moore:2000wx, Dorsch:2021nje, Friedlander:2020tnq, Cline:2020jre, Cline:2021iff, Laurent:2020gpg,Megevand:2013hwa,Dorsch:2018pat,Kozaczuk:2015owa,Konstandin:2014zta, Moore:1995si, Ai:2021kak,Lewicki:2021pgr, Ellis:2022lft, BarrosoMancha:2020fay, Liu:1992tn,Laurent:2022jrs}, here we will adopt the analytic estimate of \cite{Lewicki:2021pgr, Ellis:2022lft}
\be 
v_w =
\begin{cases}
\sqrt{\frac{\Delta V}{\alpha \rho_R}} \quad \quad {\rm for} \quad \sqrt{\frac{\Delta V}{\alpha \rho_R}}<v_J(\alpha)~,
\\
1 \quad \quad \quad \quad {\rm for} \quad  \sqrt{\frac{\Delta V}{\alpha \rho_R}} \geq v_J(\alpha) \, ,
\end{cases}
\label{eqn:approx_velocityN}
\ee
where $\a$ is the transition strength given in \eq{alphaT}
and $v_J=\frac{1}{\sqrt{3}}\frac{1+\sqrt{3 \alpha^2+2 \alpha}}{1+\alpha} $ the Chapman-Jouguet velocity which defines the upper limit for which hydrodynamic solutions can be found.
Although this result is valid for simple extensions of the SM, in our case, we expect it to give us a realistic estimate.
The reason is that we expect the friction force on the bubble wall to become significant due to the mass of the pNGBs at the metastable vacuum. 

The sound wave source template reads\footnote{We notice that there are several templates for the GW which derive from fits to different numerical simulations. In particular the template we use do not match those of, e.g.  \cite{Bringmann:2023opz} but we nevertheless expect that our conclusion remain qualitatively the same regardless of which template is used. 
} \cite{Hindmarsh:2013xza,Hindmarsh:2015qta,Hindmarsh:2017gnf,Hindmarsh:2020hop} as a function of the frequency, $f_g$,
\be\label{Omegah4}
\Omega_{\rm sw}(f_g)h^2 = 4.13\times 10^{-7} \, \left(R_* H_*\right)  \left(1- \frac{1}{\sqrt{1+2\tau_{\rm sw}H_*}} \right)  \left(\frac{\kappa_{\rm sw} \,\alpha }{1+\alpha }\right)^2 \left(\frac{100}{g_*}\right)^\frac13 S_{\rm sw}(f_g) \,,
\ee
where $R_*$ is the average bubble size at collision and the spectral function is
\be
S_{\rm sw}(f_g)=\left(\frac{f_g}{f_{\rm sw}}\right)^3 \left[\frac{4}{7}+\frac{3}{7} \left(\frac{f_g}{f_{\rm sw}}\right)^2\right]^{-\frac72},
\ee
$g_{*}$ is given in \eq{g_star_total} and all the quantities of the GW spectrum are evaluated at the nucleation temperature $T_{*}=T_n \approx T_F$. The frequency at the peak of the spectrum is given by 
\begin{equation}
 f_{\rm sw} \,=2.6\times 10^{-5}\, {\rm Hz} \left(R_* H_*\right)^{-1} \left(\frac{T_*}{100 {\rm GeV}}\right)\left(\frac{g_*}{100}\right)^\frac16 \,,
\end{equation}
while the duration of the sound wave source reads \cite{Ellis:2018mja, Ellis:2019oqb, Hindmarsh:2017gnf, Ellis:2020awk}
\begin{equation}
\tau_{\rm sw}H_* =\frac{H_* R_*}{U_f}\,, \quad U_f\approx \sqrt{\frac34 \frac{\alpha}{1+\alpha} \kappa_{\rm sw}}\,.
\end{equation}
 For the mean bubble separation we use
\begin{equation}
H_*R_* \approx (8\pi)^\frac13 \left(\frac{\beta}{H}\right)^{-1} .
\end{equation}
For all the computations  that follow in this subsection we have fixed the explicit symmetry breaking parameter to $\ve_n =10^{-2} \times \varepsilon_{n, \rm max}^{T_F}$. 
For the UV scale at which we expect new resonances to appear we have fixed $M=4\pi f$. Lastly, since the details of the dynamics of the finite temperature phase transition are to a good approximation controlled by the DS flipping temperature $T_F$, our only free parameters for this analysis are $n$, $N$, $f$, $\xi_{\rm DS}$ and $T_v$ where the visible sector temperature is fixed above $T_F$.

\begin{figure}[!ht]
\centering
\includegraphics[scale=.5]{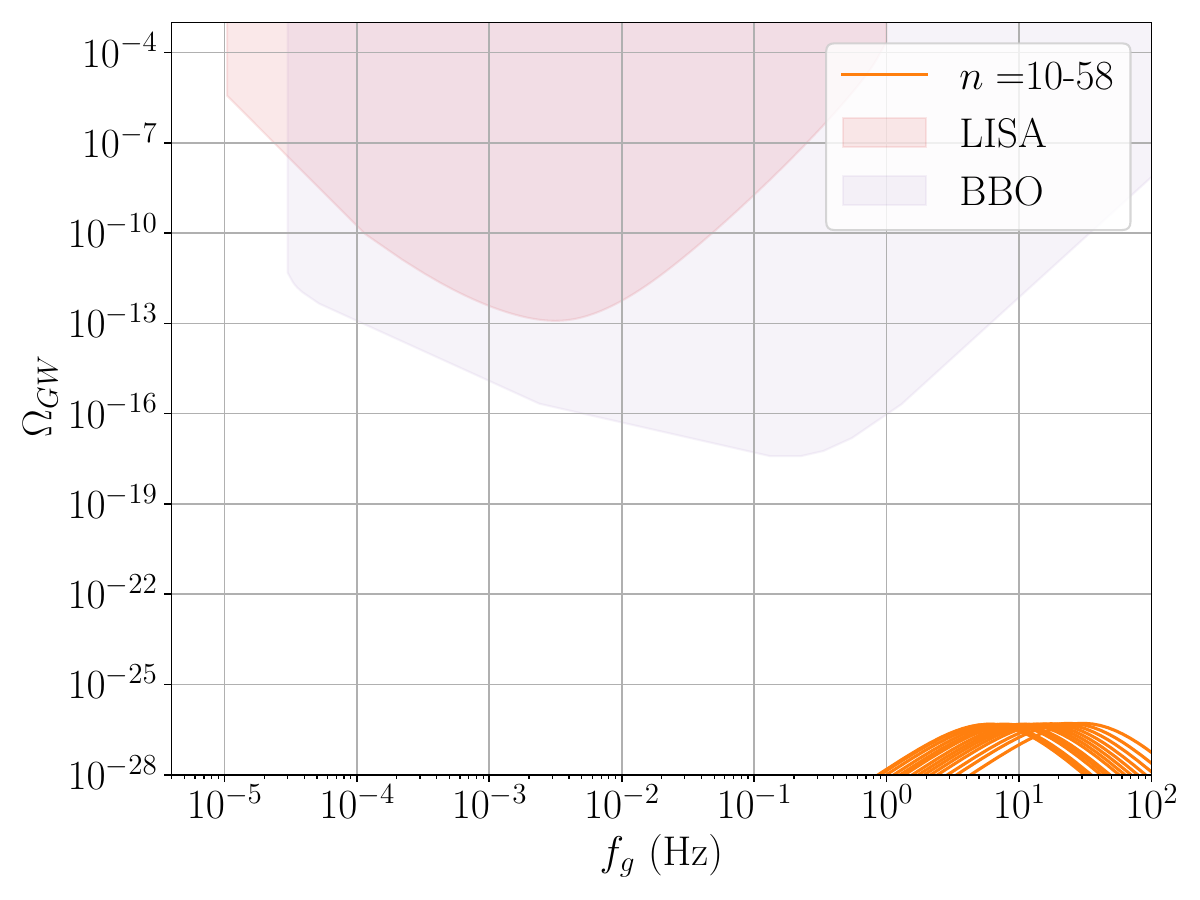} 
    \caption{\it  The GW spectrum for sound waves with  $n=10-58$ (orange curves). The explicit symmetry breaking parameter has been set to $\varepsilon=10^{-2} \times \varepsilon_{n,max}^{T_F}$. The symmetry breaking scale was fixed to $f=1$ TeV and the number of pNGBs to $N=4$. The temperature of the visible sector was fixed to $T_v = 2T_F$ for each benchmark.} \label{GW_results_1}
\end{figure}

\begin{figure}[!ht]
\centering
\includegraphics[scale=.5]{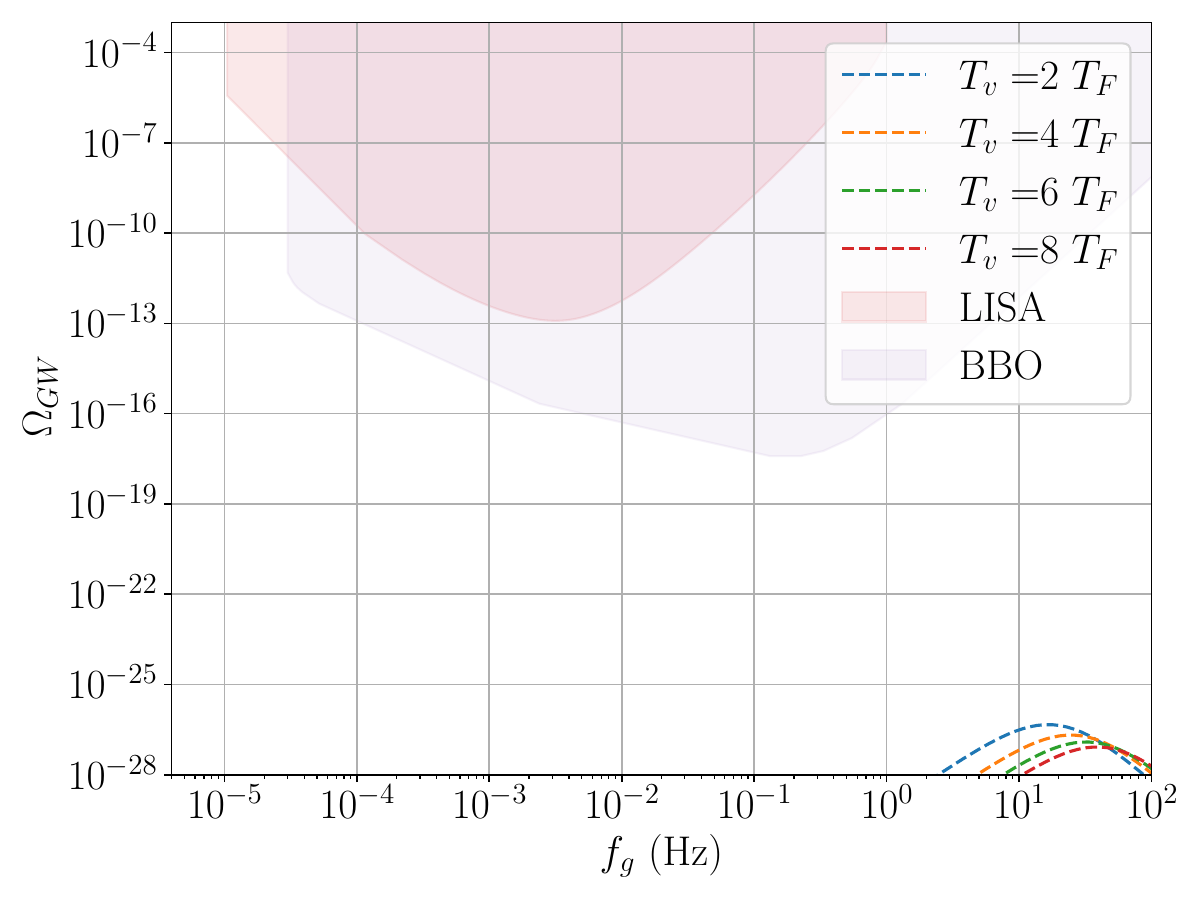} 
    \caption{\it    The GW spectrum from sound waves for  $n=20$. The explicit symmetry breaking parameter has been set to $\ve_n=10^{-2} \times \ve_{n,max}^{T_F}$. The symmetry breaking scale was fixed to $f=1$ TeV and the number of pNGBs to $N=4$. The temperature of the visible sector was fixed as specified on the plot legends.}
    \label{GW_results_2}
\end{figure}
We present the predictions of the GW spectrum in \fig{GW_results_1} and \fig{GW_results_2} below. In \fig{GW_results_1} we display the variation of the signal as a function of the Gegenabuer polynomial order, $n$, while fixing $N=4$, $f=1$ TeV and $T_v=2T_F$. We notice that the amplitude of the signal is very small compared with the expected experimental sensitivities, in particular we find $\alpha \approx 0.002$ (in agreement with our analytic prediction for the transition strength given in \eq{aTcriteTng}), $\beta/H \approx 10^6$ and $v_w \approx 0.06$. We do not observe strong dependence on the polynomial order $n$.  Recall that $T_n \approx T_F \sim f/n$, hence the flipping temperature is numerically very close to the critical and the nucleation temperature.

In \fig{GW_results_2}, we instead vary the ratio of hidden to visible temperatures by choosing different values of $T_v/T_F$ while setting $N=4$, $n=20$ and $f=1$ TeV. In this case we notice a substantial reduction in the amplitude as we increase the temperature hierarchy. This is expected as the amplitude formula \eq{Omegah4} is inversely proportional to the total number of degrees of freedom, in agreement with the results of \cite{Breitbach:2018ddu}. Furthermore we have verified numerically that varying other parameters of the potential do not substantially change the amplitude of the signal and, irrespectively of the adopted benchmark, we obtain a strength parameter of about $\alpha \approx 0.002$ while for the inverse timescale  $\beta/H \approx 10^6$ and $v_w \approx 0.06$. These numerical values are indicative of a very weak and quick transition, if not a crossover, motivating our initial choice of using $T_n$ in the GW template formula rather than the percolation temperature.

  
\bibliographystyle{JHEP}
\bibliography{references.bib}  
\end{document}